\documentclass[sigconf]{acmart}
\usepackage{adjustbox}
\usepackage{array}
\acmISBN{}
\acmDOI{}

\input{ictir25-adaptive-margins.sty}
\graphicspath{{./ictir25-adaptive-margins-figures}}

\begin{document}
\title{Learning Effective Representations for Retrieval Using Self-Distillation with Adaptive Relevance Margins}

\author{Lukas Gienapp}
\orcid{0000-0001-5707-3751}
\affiliation{
    \institution{University of Kassel \& ScaDS.AI \& hessian.AI}
    \city{Kassel}
    \country{Germany}
}

\author{Niklas Deckers}
\orcid{0000-0001-6803-1223}
\affiliation{
    \institution{University of Kassel \& hessian.AI}
    \city{Kassel}
    \country{Germany}
}

\author{Martin Potthast}
\orcid{0000-0003-2451-0665}
\affiliation{
    \institution{University of Kassel \& ScaDS.AI \& hessian.AI}
    \city{Kassel}
    \country{Germany}
}

\author{Harrisen Scells}
\orcid{0000-0001-9578-7157}
\affiliation{
    \institution{University of Tübingen}
    \city{Tübingen}
    \country{Germany}
}

\renewcommand{\shortauthors}{Lukas Gienapp, Niklas Deckers, Martin Potthast, and Harrisen Scells}

\begin{abstract}
Representation-based retrieval models, so-called bi-encoders, estimate the relevance of a document to a query by calculating the similarity of their respective embeddings. Current state-of-the-art bi-encoders are trained using an expensive training regime involving knowledge distillation from a teacher model and batch-sampling. Instead of relying on a teacher model, we contribute a novel parameter-free loss function for self-supervision that exploits the pre-trained language modeling capabilities of the encoder model as a training signal, eliminating the need for batch sampling by performing implicit hard negative mining. We investigate the capabilities of our proposed approach through extensive experiments, demonstrating that self-distillation can match the effectiveness of teacher distillation using only 13.5\% of the data, while offering a speedup in training time between 3x and 15x compared to parametrized losses.
All code and data is made openly available.\footnote{
    Code: \href{https://github.com/webis-de/adaptive-relevance-margin-loss}{github.com/webis-de/adaptive-relevance-margin-loss} \\
    \phantom{\textsuperscript{2}}Run files: \href{https://zenodo.org/records/11197962}{zenodo.org/records/11197962} \\
}

\end{abstract}

\ccsdesc[500]{Information systems~Retrieval models and ranking}
\ccsdesc[500]{Computing methodologies~Learning to rank}

\keywords{Representation Learning, Self-Distillation, Learning to Rank}


\copyrightyear{2025}
\acmYear{2025}
\setcopyright{cc}
\setcctype{by}
\acmConference[ICTIR '25]{Proceedings of the 2025 International ACM SIGIR
Conference on Innovative Concepts and Theories in Information Retrieval
(ICTIR)}{July 18, 2025}{Padua, Italy}
\acmBooktitle{Proceedings of the 2025 International ACM SIGIR Conference on
Innovative Concepts and Theories in Information Retrieval (ICTIR) (ICTIR '25), July
18, 2025, Padua, Italy}\acmDOI{10.1145/3731120.3744594}
\acmISBN{979-8-4007-1861-8/2025/07}




\makeatletter
\gdef\@copyrightpermission{
 \begin{minipage}{0.3\columnwidth}
  \href{https://creativecommons.org/licenses/by/4.0/}{\includegraphics[width=0.90\textwidth]{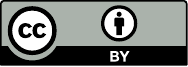}}
 \end{minipage}\hfill
 \begin{minipage}{0.7\columnwidth}
  \href{https://creativecommons.org/licenses/by/4.0/}{This work is licensed under a Creative Commons Attribution International 4.0 License.}
 \end{minipage}
 \vspace{5pt}
}
\makeatother

\maketitle
\section{Introduction}

In information retrieval, trans\-for\-mer-based bi-encoders are used as effective retrieval models. They estimate a document's relevance to a query by computing the similarity of their embeddings. Queries and documents are represented independently, which allows for pre-computing and indexing all document representations offline and computing only the query representation at retrieval time. Bi-encoders are thus appealing in practice since they can be easily scaled. In contrast, transformer-based cross-encoders are a type of model that estimates relevance jointly by computing a score for each document--query pair at retrieval time, achieving higher effectiveness than bi-encoders at the expense of a higher computational cost and inference latency. They are thus usually reserved for multi-stage re-ranking, such that only the top-$k$ documents of an initial ranking produced by a more efficient bi-encoder are re-ranked.

\begin{figure}[t]
    \includegraphics[width=\linewidth,trim={0cm .175cm 0cm 0cm},clip]{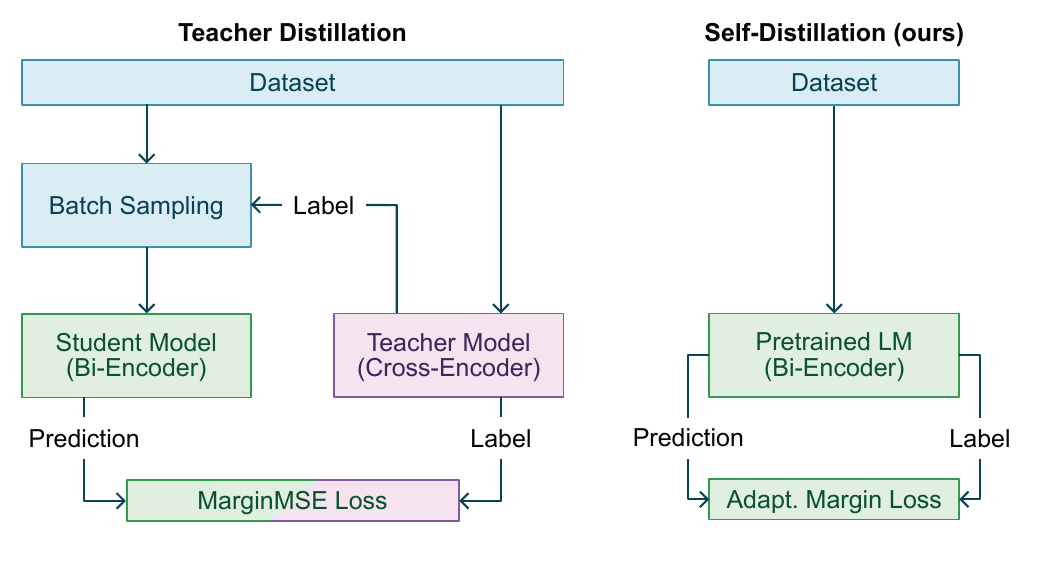}
    \caption{Conceptual overview of a typical bi-encoder teacher distillation training regime (left) and our proposed self-distillation approach (right). Blue is the data, green is the bi-encoder and purple is the cross-encoder model.}
    \label{fig:conceptual-overview}
\end{figure}

However, highly effective bi-encoders come at the cost of a highly expensive training regime based on knowledge distillation. Figure~\ref{fig:conceptual-overview} (left) illustrates this process, where the relevance of each training sample is first estimated by a cross-encoder teacher model. The estimated relevance scores are first used for a computationally intensive batch sampling process, where an optimal distribution of relevance scores within each batch is ensured according to the teacher model, which then provides supervision for training a bi-encoder student model with a margin-based loss function. While this knowledge distillation setup yields highly effective bi-encoder ranking models, we highlight three drawbacks: 
\begin{enumerate*}[label=\bfseries(\arabic*)]
\item 
A teacher model must be available for the desired dataset and domain. Cross-encoder models are usually chosen as teacher models to maximize effectiveness~\cite{hofstatter:2020}. Training a cross-encoder model for knowledge distillation is computationally expensive and may not be possible for domains with little ground-truth ranking data.
\item 
Teacher scores must be inferred for every training sample, requiring a forward pass. This adds inference cost onto a process that is already prohibitively expensive. Improving training efficiency is crucial for information retrieval, as the majority of energy spent over the lifetime of a model, especially models that are efficient at retrieval time, goes into training~\cite{scells:2022}.
\item 
Batch sampling prohibits continuous learning on new data. The current best batch sampling procedures rely on clustering the training data set~\cite{hofstatter:2020}. A static training dataset is thus required, rendering it challenging to adapt such approaches in settings where training data is collected continuously, such as online learning-to-rank.
\end{enumerate*}

In this paper, after a careful review of related work (Section~\ref{sec:related}), we propose a novel training regime and loss function for bi-encoders (Figure~\ref{fig:conceptual-overview}, right) that simplifies previous approaches while maintaining competitive effectiveness (Section~\ref{sec:method}). Instead of using a teacher model, we leverage the pre-trained text similarity capabilities of the encoder model to provide a supervision signal in a self-distillation setup. This eliminates the need for expensive exhaustive inference over the training dataset, does not rely on batch sampling techniques, and is highly data efficient. Furthermore, our proposed loss function is free of hyperparameters. We then conduct systematic experiments (Section~\ref{sec:setup}) which demonstrate that self-distillation can match and even surpass the effectiveness of teacher distillation-based training regimes in in-domain settings, and still sufficiently generalize to zero-shot ranking tasks, while using only~13.5\% of the data (Section~\ref{sec:experiment}).
\section{Preliminaries \& Related Work}\label{sec:related}

We define $Q$ as a set of queries representing user information needs and~$D$ as a set of documents potentially containing relevant information. A retrieval model~$\rho: Q\times D \rightarrow \mathbb{R}$ maps pairs of a document $d \in D$ and a query~$q\in Q$ to a relevance score, where a higher score indicates higher relevance of~$d$ to~$q$. We summarize related work on operationalizing $\rho$ with representation learning, covering encoder models (\Cref{sec:encoder-models}), knowledge distillation (\Cref{sec:knowledge-distillation}), loss functions (\Cref{sec:loss-functions}), and negative mining (\Cref{sec:negative-mining}).

\subsection{Relevance Estimation with Encoder Models}\label{sec:encoder-models}

Text encoder models like BERT~\cite{devlin:2019} operationalize~$\rho$ in two ways: bi-encoders and cross-encoders~\cite{zhao:2022}. Bi-encoders independently embed the query~$q$ and each document~$d$ using an encoder model~$f_\theta$, typically by using the representation of the special \texttt{[CLS]} token (optionally processed by a linear layer)~\cite{chang:2020,gao:2021a}. They measure relevance in terms of a similarity function~$\varphi$ as $\rho = \varphi(f_\theta(q), f_\theta(d))$~\cite{karpukhin:2020,chang:2020,gao:2021b}. In contrast, cross-encoders jointly process query and document, operationalizing~$\rho$ as a relevance classification of the joint representation~\cite{nogueira:2019}: $\rho = c(f_\theta(q \oplus d))$, where~$\oplus$ is string concatenation and $c: \mathbb{R}^D\rightarrow [0,1]$ is a classifier assigning a probability of relevance. While bi-encoders are more efficient~\cite{zhao:2022}, cross-encoders capture complex interactions between query and document~\cite{nogueira:2019}, yielding higher effectiveness albeit with higher latency at query time, as document representations cannot be pre-computed. We focus our investigation on training bi-encoders with comparable effectiveness to cross-encoders.

\subsection{Knowledge Distillation}\label{sec:knowledge-distillation}

Knowledge distillation~\cite{hinton:2015} aims to train a student model to reproduce the output of a teacher model~\cite{hu:2023}. It is commonly employed for model compression, i.e., to develop smaller and faster models capable of maintaining effectiveness at reduced computational effort~\cite{gou:2021}. In retrieval tasks, distillation, as proposed by~\citet{hofstatter:2020,hofstatter:2021}, trains bi-encoder models using pre-computed relevance scores from a cross-encoder teacher model. Although distillation can successfully impart ranking effectiveness to bi-encoders, the use of cross-encoder teachers imposes constraints on the training process due to the computational cost of  pre-computing relevance scores for a large set of query--document pairs. Thus, while distillation produces highly efficient student models, it incurs high computational costs for training.

Self-distillation is a subfield of knowledge distillation, where the student and teacher models are identical~\cite{hahn:2019}, i.e., leveraging a model's own output to provide a training signal for a refined version of itself. The success of this approach is attributed to the rich output distribution, capturing additional information about training examples~\cite{liu:2021}. While self-distillation is predominantly applied when the teacher and student models are trained towards the same task, it can also be used in a transfer setting~\cite{zhou:2022}. Here, the pre-trained general capabilities of the model are exploited to train a more refined version of itself on a specialized target task.This reduces the required computational effort, as no additional teacher inference is needed. We take inspiration from prior work in the literature about self-distillation to apply to the task of ranking.

\subsection{Bi-Encoder Loss Functions}\label{sec:loss-functions}

In the typical training setup for bi-encoder models, data comes in the form of triplets~$(q, d^+, d^-)$, comprising a query, a positive document~$d^+$, and a negative document~$d^-$. Various loss functions have been proposed within this setup: for instance, triplet loss aims to adjust the representation space so that the positive document~$d^+$ is closer (i.e., more relevant) to the query~$q$ than the negative document~$d^-$ by a certain margin~$\varepsilon$~\citep{schroff:2015}. \citet{karpukhin:2020} minimize the negative log-likelihood between a single positive document and multiple negative documents. \citet{xiong:2021} minimize a classification loss function like binary cross-entropy, hinge loss, or negative log-likelihood between scores predicted for documents pairs. All aim to rearrange representations to prioritize relevance, i.e., to learn a representation space in which for a query, a relevant document is closer than a non-relevant document by a margin~$\varepsilon$.

In knowledge distillation contexts, teacher scores provide a source of supervision. The MarginMSE loss~\cite{hofstatter:2020} adopts the concept of optimizing the mean squared error between the relevance difference, i.e., $\varphi(q,d^+) - \varphi(q, d^-)$, and a desired target~$\varepsilon$. The value of~$\varepsilon$ is dictated individually for each triplet by its predicted teacher model output. This notion of introducing an `adaptive' target value for each instance also has broader applications beyond distillation. For instance, \citet{li:2020} use class distances in an image classification problem as adaptive target, \citet{ha:2021} use differences in ground-truth labels as adaptive targets for learning-to-rank, and \citet{sharma:2024} use the similarity between augmented views of the same sample. We apply self-distillation to predict an adaptive target for each training triplet using the bi-encoder model itself. 

\subsection{In-Batch Negatives \& Hard Negative Mining}\label{sec:negative-mining}

To increase the efficiency of triplet-based losses, additional training instances can use other triplets in the same batch as supplementary negative documents for a given pair of query and positive document~\cite{zhan:2021}. A major caveat is that~$f_\theta$ rapidly learns to correctly project trivial triplets, i.e., triplets where~$d^+$ and~$d^-$ are easy to distinguish with respect to their relevance. Consequently, identifying `hard' triplets, where distinguishing between positive and negative documents is challenging, becomes crucial for learning a robust latent space~\cite{zhan:2021}. Yet, in-batch negatives are likely topically unrelated to a given query, and thus have limited learning potential.

To alleviate this issue, batch sampling techniques have been devised that select batches of triplets such that the positive and negative documents are hard to distinguish in terms of their relevance to~$q$, both within each pair and across pairs of in-batch negatives~\cite{cai:2022}. For retrieval, \citet{hofstatter:2021} introduce balanced topic-aware sampling~(TAS-B), where triplets are chosen to ensure that the relevance differences of query--positive and query--negative pairs across all training pairs in a batch adhere to a certain distribution. All triplets in the training data are put into bins based on their margin, and then batches are constructed by uniformly selecting triplets from all bins. A reproduction study by \citet{wang:2023} found that effectiveness can be further improved by discarding bins with high relevance differences, removing `easy' negatives from consideration. By predicting adaptive targets, we show that self-distillation allows `implicit' negative sampling based on the magnitude of the target without explicitly sampling batches.

  \section{Self-Distillation With Adaptive Relevance Margins}\label{sec:method}

We formulate three criteria to simplify the training process, increase the training efficiency of bi-encoders for retrieval, and address the drawbacks of teacher distillation approaches outlined before:
\begin{enumerate*}[label=(\arabic*)]
\item 
no teacher model for target estimation: the loss function should only rely on the information contained in provided training triplets;
\item 
no batch sampling procedure: the loss function should be effective with randomly ordered data;
\item 
no hyperparameters: the loss function should be applicable without hyperparameter tuning.
\end{enumerate*}
In this section, we introduce \emph{adaptive relevance margins}, a novel self-distillation loss function for retrieval fine-tuning. We motivate it starting from how a margin loss is traditionally calculated, and derive several variants for the inner term of the loss.

\paragraph{Static Targets.}
The traditional margin loss for the $i$-th training triplet $(q_i, d^+_i, d^-_i)$ consisting of a query~$q_i$, a positive (i.e., relevant) document~$d^+_i$, and a negative (i.e., less or non-relevant) document~$d^-_i$ contrasts the query--document relevance margin~\cite{hofstatter:2021}. It is calculated as the difference between the cosine similarity~$\varphi(q_i, d^+_i)$ of~$q_i$ and~$d^+_i$ and the cosine similarity~$\varphi(q_i, d^-_i)$ of~$q_i$ and~$d^-_i$, with a target value~$\varepsilon\in[0,1]$:%
\footnote{For brevity, we omit the encoder~$f_\theta$ from the loss functions.}

\begin{equation}
\label{eq:marginmse}
    l_{i} = 
    \underbrace{
       \varphi\left(q_i, d^+_i\right) 
        - \varphi\left(q_i, d^-_i\right)
    }_{\text{Relevance margin}}
    - \kern-.75em
    \underbrace{
        \vphantom{\left(\varphi(d^+_i, d^-_i)\right)} 
        \varepsilon
    }_{\text{Target}}
\end{equation}

This loss optimizes for a latent space where positive and negative documents are separated by a value of at least~$\varepsilon$ in terms of their similarity to 
a query, where $\varepsilon$ is a hyperparameter to be tuned.

\paragraph{Adaptive Targets.}
Rather than using the same $\varepsilon$ value as a target across all training instances, knowledge distillation allows one to derive per-document target scores using a teacher model~\cite{hofstatter:2021}. We propose using the bi-encoder model itself as a source for this teacher signal through self-distillation. As BERT-style encoders are pre-trained with a masked token or sentence prediction objective, they exhibit strong capabilities for measuring text similarity~\cite{devlin:2019,liu:2019}. We can, thus, replace~$\varepsilon$ with the normalized similarity between positive and negative documents for a query, scaled to the $[0,1]$~range:

\begin{equation}
\label{eq:adaptive}
    l_{i} = 
    \underbrace{
        \vphantom{\left(\frac{1 + \varphi(d^+_i, d^-_i)}{2}\right)} 
        \varphi\left(q_i, d^+_i\right) 
        - \varphi\left(q_i, d^-_i\right)
    }_{\text{Relevance margin}}
    - 
    \underbrace{
        \left(\frac{1 + \varphi(d^+_i, d^-_i)}{2}\right)
    }_{\text{\begin{tabular}[t]{c}
      Scaled document\\[-1ex]
      similarity
      \end{tabular}}}
\end{equation}

Rescaling the target is warranted, since we desire a latent space where positive and negative documents are orthogonal. In particular, we designed this adaptive margin to learn to predict a larger difference in relevance for similar documents rather than dissimilar ones. If the positive and negative documents are similar (i.e., $(1 + \varphi(d^+, d^-)) / 2 \approx 1$), they are a `hard' training instance, which results in a higher contribution to the overall loss and therefore a higher impact on the model parameters compared to dissimilar documents (i.e., $(1 + \varphi(d^+, d^-)) / 2 \approx 0$). Adaptive targets thus implicitly perform hard negative mining without hyperparameters.

\paragraph{Distributed Targets.}
So far, the loss calculation has used only a single triplet~$(q_i, d_i^+, d_i^-)$. While na\"ive in-batch negative variants~$l_{ij}$ of the static and adaptive targets losses can be derived by replacing~$d_i^-$ with multiple $d_j^- \in D^-$, this yields sub-par results if not combined with batch sampling techniques. Instead, we extend the adaptive targets approach to approximate this sampling by using multiple different target values for each individual training triplet:

\begin{equation}
\label{eq:distributed}
    l_{ij} = 
        \underbrace{
            \vphantom{\left(\frac{1 + \varphi(d^+_i, d^-_j)}{2}\right)}
            \varphi\left(q_i, d^+_i\right) - \varphi\left(q_i, d^-_i\right) 
        }_{\text{Relevance margin}} - 
        \underbrace{
            \left(\frac{1 + \varphi(d^+_i, d^-_j)}{2}\right) 
        }_{\text{\begin{tabular}[t]{c}
          Scaled document\\[-1ex]
          similarity distribution
          \end{tabular}}}
\end{equation}

The intuition behind this approach is that the target now forms a distribution over the negative documents~$D^-$ in the batch. Instead of comparing each triplet in the batch with a single target, we repeatedly compare the margin to the original negative with targets from all of the negatives in a batch. In other words, these repeated comparisons estimate the average scaled similarity of a positive document~$d^+$ to (a sample of) the population of negative documents~$D^-$. If the similarity is high~($\approx 1$), the sample contains documents that are hard to distinguish in terms of relevance and thus comprises rich information to establish a positive--negative class separation. If the similarity is low ($\approx 0$), the sample contains documents that are easy to distinguish in terms of relevance. A high in-batch negative document similarity translates to a large margin to maximize the impact on the loss, and vice versa. 

While adaptive targets opt to use in-batch information to increase the count of total available training triplets, distributed targets instead exploit the in-batch distribution to make a more accurate estimation of the target for each original triplet. One can also consider the distributed targets approach as optimizing for the implicit mean document distance. The reason optimizing it implicitly works and optimizing it explicitly (i.e., by taking the mean scaled document similarity within the loss as target) does not work is because optimizing it explicitly repositions only~$d_i^-$ in latent space rather than all $d_j^-\in D^-$ as the implicit mean does.

\paragraph{Error Function.}
To form a proper loss function, the previously derived pointwise terms~$l_{ij}$ are optimized using mean squared error~(MSE), which calculates the average squared difference of the relevance margin and target over all samples in a batch. A batch consists of a set of queries~$Q$, a set of associated positive (relevant) documents~$D^+$, and a set of associated negative (less or non-relevant) documents~$D^-$, with batch size~$B = |Q| = |D^+| = |D^-|$. For all variants~$l_i$ not relying on in-batch information, the error is calculated as

\begin{equation}
    \mathcal{L}_{MSE}\left(Q, D^+\kern-.25em,D^-\right) = \frac{1}{B}\sum_{i=1}^B\left(l_{i}\right)^2
\end{equation}

\noindent
For variants~$l_{ij}$ using in-batch information it is calculated as

\begin{equation}
    \mathcal{L}_{MSE}\left(Q, D^+\kern-.25em,D^-\right) = \frac{1}{B^2}\sum_{i=1}^B\sum_{j=1}^B \left(l_{ij}\right)^2
\end{equation}

\begin{figure}[t]
    \hbox to \linewidth{
    \vbox{
        \hbox to .3\linewidth{{\small\textbf{(a)} Static}\hfill}
        \hbox to .3\linewidth{\hfill
            \begin{tikzpicture}[every node/.style = {fill, circle, inner sep=1pt, font=\small}]
                \coordinate (root) at (0,0);
                \draw[name path=unit, dotted] (root) circle [radius=.75];
                \node[label={90:$q$}] (q) at (xyz polar cs:angle=90,radius=.75) {};
                \node[label={45:$d^+$}] (p) at (xyz polar cs:angle=45,radius=.75) {};
                \node[label={143:$d^-_2$}] (n1) at (xyz polar cs:angle=143,radius=.75) {};
                \node[label={-15:$d^-_1$}] (n2) at (xyz polar cs:angle=-15,radius=.75) {};
                \node[label={-96:$d^-_3$}] (n3) at (xyz polar cs:angle=-96,radius=.75) {};
                
                \coordinate (coneA) at (xyz polar cs:angle=-45,radius=.75);
                \coordinate (coneB) at (xyz polar cs:angle=-135,radius=.75);
                \path[draw=Green3, ultra thick, opacity=0.2] (coneA) --  (root) -- (coneB);
                \path (coneB) -- (root) -- (n1) pic [-Latex, thick, draw=Red3, angle radius=.75cm, fill=none] {angle = n1--root--coneB};
                \path (n2) -- (root) -- (coneA) pic [Latex-, thick, draw=Red3, angle radius=.75cm, fill=none] {angle = coneA--root--n2};
                \path (coneA) -- (root) -- (n3) pic [Latex-, thick, draw=Red3, angle radius=.75cm, fill=none] {angle = coneB--root--n3};
            \end{tikzpicture}
        \hfill}
    }
    {\textcolor{Gray3}\vrule}
    \hfill
    \vbox{
        \hbox to .3\linewidth{{\small\textbf{(b)} Adaptive}\hfill}
        \hbox to .3\linewidth{\hfill
            \begin{tikzpicture}[every node/.style = {fill, circle, inner sep=1pt, font=\small}]
                \coordinate (root) at (0,0);
                \draw[name path=unit, dotted] (root) circle [radius=.75];
                \node[label={90:$q$}] (q) at (xyz polar cs:angle=90,radius=.75) {};
                \node[label={45:$d^+$}] (p) at (xyz polar cs:angle=45,radius=.75) {};
                \node[label={143:$d^-_2$}] (n1) at (xyz polar cs:angle=143,radius=.75) {};
                \node[label={-15:$d^-_1$}] (n2) at (xyz polar cs:angle=-15,radius=.75) {};
                \node[label={-96:$d^-_3$}] (n3) at (xyz polar cs:angle=-96,radius=.75) {};
                
                \node[fill=Green3, inner sep=3pt, opacity=0.2] (x1) at (xyz polar cs:angle=193,radius=.75) {};
                \node[fill=Green3, inner sep=3pt, opacity=0.2] (x2) at (xyz polar cs:angle=10,radius=.75) {};
                \node[fill=Green3, inner sep=3pt, opacity=0.2] (x3) at (xyz polar cs:angle=-126,radius=.75) {};
                
                \path (x1) -- (root) -- (n1) pic [-Latex, thick, draw=Red3, angle radius=.75cm, fill=none] {angle = n1--root--x1};
                \path (x2) -- (root) -- (n2) pic [-Latex, thick, draw=Red3, angle radius=.75cm, fill=none] {angle = n2--root--x2};
                \path (x3) -- (root) -- (n3) pic [Latex-, thick, draw=Red3, angle radius=.75cm, fill=none] {angle = x3--root--n3};
            \end{tikzpicture}
        \hfill}
    }
    {\textcolor{Gray3}\vrule}
    \hfill
    \vbox{
        \hbox to .3\linewidth{{\small\textbf{(c)} Distributed}\hfill}
        \hbox to .3\linewidth{\hfill
            \begin{tikzpicture}[every node/.style = {fill, circle, inner sep=1pt, font=\small}]
                \coordinate (root) at (0,0);
                \draw[name path=unit, dotted] (root) circle [radius=.75cm];
                \node[label={90:$q$}] (q) at (xyz polar cs:angle=90,radius=.75) {};
                \node[label={45:$d^+$}] (p) at (xyz polar cs:angle=45,radius=.75) {};
                \node[Gray3,label={[Gray3]143:$d^-_2$}] (n1) at (xyz polar cs:angle=143,radius=.75) {};
                \node[label={[font=\bfseries]-15:$d^-_1$}] (n2) at (xyz polar cs:angle=-15,radius=.75) {};
                \node[Gray3,label={[Gray3]-96:$d^-_3$}] (n3) at (xyz polar cs:angle=-96,radius=.75) {};

                \coordinate (x11) at (xyz polar cs:angle=-37,radius=.75);
                \coordinate (x21) at (xyz polar cs:angle=217,radius=.75);
                \path (x11) -- (root) -- (x21) pic [fill=Green3, angle radius=.5cm, opacity=0.15] {angle = x21--root--x11};
                
                \coordinate (x12) at (xyz polar cs:angle=-18,radius=.75);
                \coordinate (x22) at (xyz polar cs:angle=198,radius=.75);
                \path (x22) -- (root) -- (x12) pic [fill=Green3, angle radius=.5cm, opacity=0.15] {angle = x22--root--x12};

                \coordinate (x13) at (xyz polar cs:angle=9,radius=.75);
                \coordinate (x23) at (xyz polar cs:angle=170,radius=.75);
                \path (x13) -- (root) -- (x23) pic [fill=Green3, angle radius=.5cm, opacity=0.15] {angle = x23--root--x13};
                
                \coordinate (t) at (xyz polar cs:angle=-55,radius=.75);
                \node[fill=Green3, inner sep=3pt, opacity=0.2] at (t) {};
                \path (x1) -- (root) -- (n2) pic [Latex-, thick, draw=Red3, angle radius=.75cm, fill=none] {angle = t--root--n2}; 
            \end{tikzpicture}
        \hfill}
    }
}
    \caption{Conceptual illustration of target variants in cosine similarity space.  Black dots indicate location on the unit sphere. Green elements indicate target values; red arrows indicate induced shifts in latent space for negative documents.}
    \label{fig:loss-variants}
\end{figure}

\paragraph{Discussion.}
Figure~\ref{fig:loss-variants} illustrates the behavior of all three loss variants. The position of a query~$q$, a positive document~$d^+$, and a set of negative documents~$d^-_i$ are visualized in cosine similarity space. Green elements indicate the target values and red arrows indicate the repositioning in latent space induced by the respective loss. We only illustrate repositioning of negative documents, for simplicity; in practice, all three, query, positive, and negative could be simultaneously affected. For
\begin{enumerate*}[label=(\alph*)]
\item
static targets, a relevance margin of $\varepsilon$ is forced (green lines); for
\item
adaptive targets, individual relevance margins for each negative are estimated based on document similarity (green dots); and for
\item
distributed targets, for a single negative~$d^-_1$, the distribution of the document similarities of all negatives (green area) is used as margin (green dot).
\end{enumerate*}

\section{Experimental Setup}\label{sec:setup}

\paragraph{Data.}
We employ the MSMARCO-Passage collection~\cite{nguyen:2016}. For training, we use the official training set of~\SI{39780811}{} training triplets, each comprising a query, positive, and negative document (i.e., text passage in this case). For evaluation, we employ relevance annotations from the TREC-DL~19~\cite{craswell:2019} and TREC-DL~20~\cite{craswell:2020} datasets, which contain~43 and~54 densely judged queries for the MSMARCO-Passage collection, respectively. Since TREC annotations include graded relevance (ranging from~0 for non-relevant to~3 for maximum relevance), we use the recommended binarization level of $r > 1$~\cite{craswell:2019} for all binary metrics. Text is tokenized with a maximum sequence length of 30 for queries and 200 for passages~\cite{hofstatter:2021}.

\paragraph{Implementation and Hyperparameters.}
We implement our experiment pipeline in PyTorch~\cite{paszke:2019} using the Lightning~\cite{falcon:2019} and Hugging Face~\cite{wolf:2020} libraries. We fine-tune several pre-trained BERT-style encoder models, which vary in parameter count, pre-training regime, and embedding kind. The full list is provided in \autoref{tab:models}. Text representations are computed using the \texttt{[CLS]} representation of each document, optionally processed by a linear layer. All models produce 768-dimensional embeddings. We train using the ADAM optimizer with a learning rate of $5e^{-6} / B$, a weight decay rate of $1e^{-6}$, and batch sizes $B\in[ 64, 128]$. 

\paragraph{Hardware.}
Training is performed on a single Nvidia A100 40GB GPU, which renders our training setup attainable for most academic and small-scale industrial research environments. To improve training efficiency, we use mixed-precision weights and quantized optimizer states, reducing memory footprint by approximately 50\% compared to full-precision training while maintaining comparable model effectiveness. 

\paragraph{Evaluation.}
We study two evaluation setups: re-ranking and full ranking. In the re-ranking setup, which is used for evaluating runs with low overhead, we re-rank the top-1000 documents from the official TREC-DL~19/20 BM25 baseline runs for each query. For the full ranking setup, replicating real-world use cases, we embed and index the entire MSMARCO-Passage collection using a FAISS~\cite{johnson:2019} HNSW index with optimized 32-byte product quantization, and perform exact re-ranking of the top-$k$ results. In both setups, we use cosine similarity between query and document representations for relevance estimation. 
Similar to prior work, in the full ranking setup we report the measures nDCG@10, representing ranking-oriented tasks, and Recall@1000, representing retrieval-oriented tasks. However, in the re-ranking setup, the BM25 baseline run artificially restricts the recall obtainable. Therefore, we use Hits@100 as a suitable surrogate for recall that does not rely on all relevant documents being retrievable for each topic.

\paragraph{Baselines.} To contextualize the results of our proposed loss functions, we compare against three different baseline models: 
\begin{enumerate*}[label=\bfseries(\arabic*)]
\item 
BM25 given its nature as standard retrieval baseline and re-ranking input in our re-ranked evaluation setting;
\item 
a BERT-based bi-encoder fine-tuned with a static margin of $\varepsilon=1$~\cite{hofstatter:2021}
\item 
a BERT-based bi-encoder fine-tuned with TAS-B batch sampling and teacher margins~\cite{hofstatter:2021,wang:2023}. 
\end{enumerate*}


\begin{table}[t]
    \caption{Overview of pre-trained models used for fine-tuning.}
    \label{tab:models}
    \centering 
    \setlength{\tabcolsep}{7.5pt}
\renewcommand{\arraystretch}{.8}
\small
\begin{tabular}{@{}lllS[table-format=3.0]lS[table-format=3.0]@{}}
\toprule
\bf Model  &                                               &                   & {\bf \textnumero~Params}       & \bf Embedding             & {\bf \textnumero~Dim.} \\ \midrule
RoBERTa    & \hflink{FacebookAI/roberta-base}              & \cite{liu:2019}   & 124 {M}               & \texttt{[CLS]} + Linear   & 768  \\
MPNet      & \hflink{microsoft/mpnet-base}                 & \cite{song:2020}  & 109 {M}               & \texttt{[CLS]} + Linear   & 768  \\
distilBERT & \hflink{distilbert/distilbert-base-uncased}   & \cite{sanh:2019}  &  66 {M}               & \texttt{[CLS]}            & 768  \\
\bottomrule
\end{tabular}

\end{table}
\section{Results \& Discussion}\label{sec:experiment}

We conduct four experiments to study the benefits of using adaptive and distributed margins for fine-tuning bi-encoder retrieval models. First, we conduct an experiment on static margins to see how adjusting the parameter~$\varepsilon$ affects effectiveness, with and without in-batch negatives (\Cref{sec:exp-static}). Second, we study the effectiveness of adaptive margins, including experiments on batch size and ablating in-batch negatives (\Cref{sec:exp-adaptive}). Third, we repeat this experiment for distributed margins (\Cref{sec:exp-distributed}). Finally, we compare the optimal parameter settings of each loss variant found in the previous experiments (\Cref{sec:exp-comparison}), and compare them to other state-of-the-art and baseline models (\Cref{sec:exp-baselines}). This comparison provides insights into the overall effectiveness of our proposed approach.

\subsection{Static Targets}\label{sec:exp-static}

In the first experiment, we conduct a grid search to find the optimal static target hyperparameter~$\varepsilon$. We fine-tune each of the three pre-trained encoders using the static target loss variant, while varying~$\varepsilon$ in the range of~0 to~1 in steps of~0.1. To make this grid search feasible in terms of computation time, we train each configuration on the same randomly sampled subset of MSMARCO-Passage, containing \SI{2.56}{M~triplets}. Evaluation scores are calculated in the re-ranking setting. The maximum batch size possible across all models on the available hardware was used~($B=128$). Experiments with smaller batch size~($B=64$) showed the same trends and relative differences. 

\begin{figure}[t]
    \includegraphics[width=\linewidth,trim={0 .7cm 0 .3cm},clip]{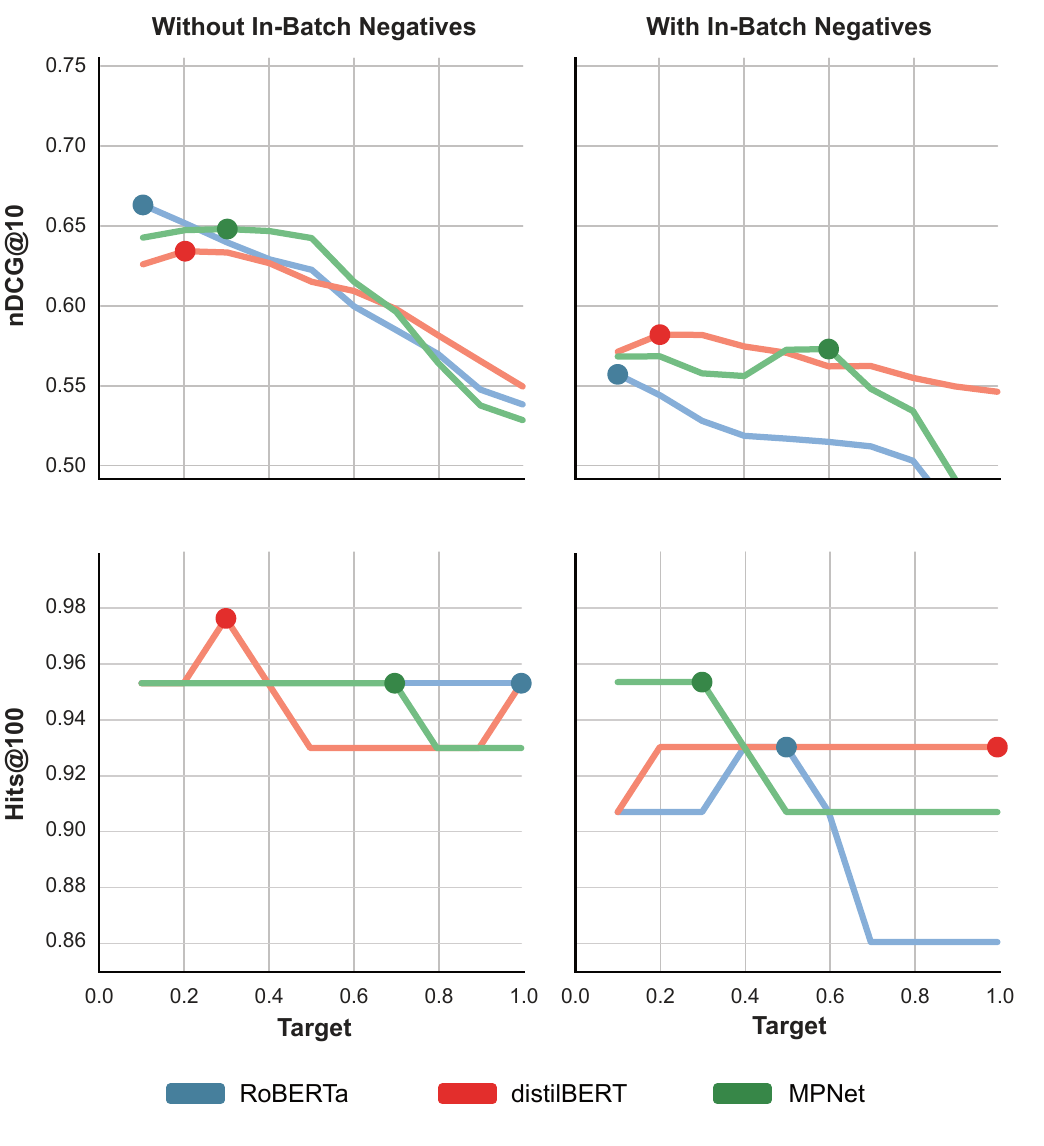}
    \caption{nDCG@10 (upper row) and Hits@100 scores (lower row) for different static target values on TREC-DL~19, without (left) and with in-batch negatives (left). Dots mark maxima.}
    \label{fig:static-margins}
\end{figure}

\autoref{fig:static-margins} shows evaluation scores for each combination of~$\varepsilon$ and model, with and without using in-batch negatives. Four effects are apparent: 
\begin{enumerate*}[label=(\arabic*)]
\item
Tuning $\varepsilon$ as a hyperparameter yields improvements across all configurations compared to the default choice of $\varepsilon=1$. Moreover, the improvement between the default and optimal configuration is distinct, surpassing 0.1 nDCG@10 in almost all cases. Consequently, margin optimization is necessary to extract maximum effectiveness in each setup.
\item
The optimal value of $\varepsilon$ varies depending on the model; each model achieves its highest evaluation score at different points. This indicates that hyperparameter optimization needs to be repeated for each pre-trained model, and there is no universal choice for retrieval fine-tuning as a whole.
\item
The optimal margin varies depending on the task; all tested models show a different $\varepsilon$-value at which the respective evaluation measure, i.e., nDCG@10 or Hits@100, is maximized.
\item 
Using in-batch negatives does not yield an improvement in effectiveness. For all three tested models and both measures, the effectiveness without in-batch negatives is the same or higher than with in-batch negatives. This suggests that the models do not benefit from the additional training data, likely because it is noisy and provides little information, as negatives in randomly sampled batches are `easy'.
\end{enumerate*}

The effectiveness of the static target loss variant, whether with or without in-batch negatives, is highly susceptible to the training setup used. Factors such as model choice, value of $\varepsilon$ result in varying and interdependent levels of effectiveness. Substantial gains in effectiveness can be achieved through hyperparameter tuning. However, negative mining techniques beyond simply random sampling appear necessary to effectively use in-batch information.

\subsection{Adaptive Targets}\label{sec:exp-adaptive}

\begin{figure}[t]
    \includegraphics[width=\linewidth,trim={0 .7cm 0 .3cm},clip]{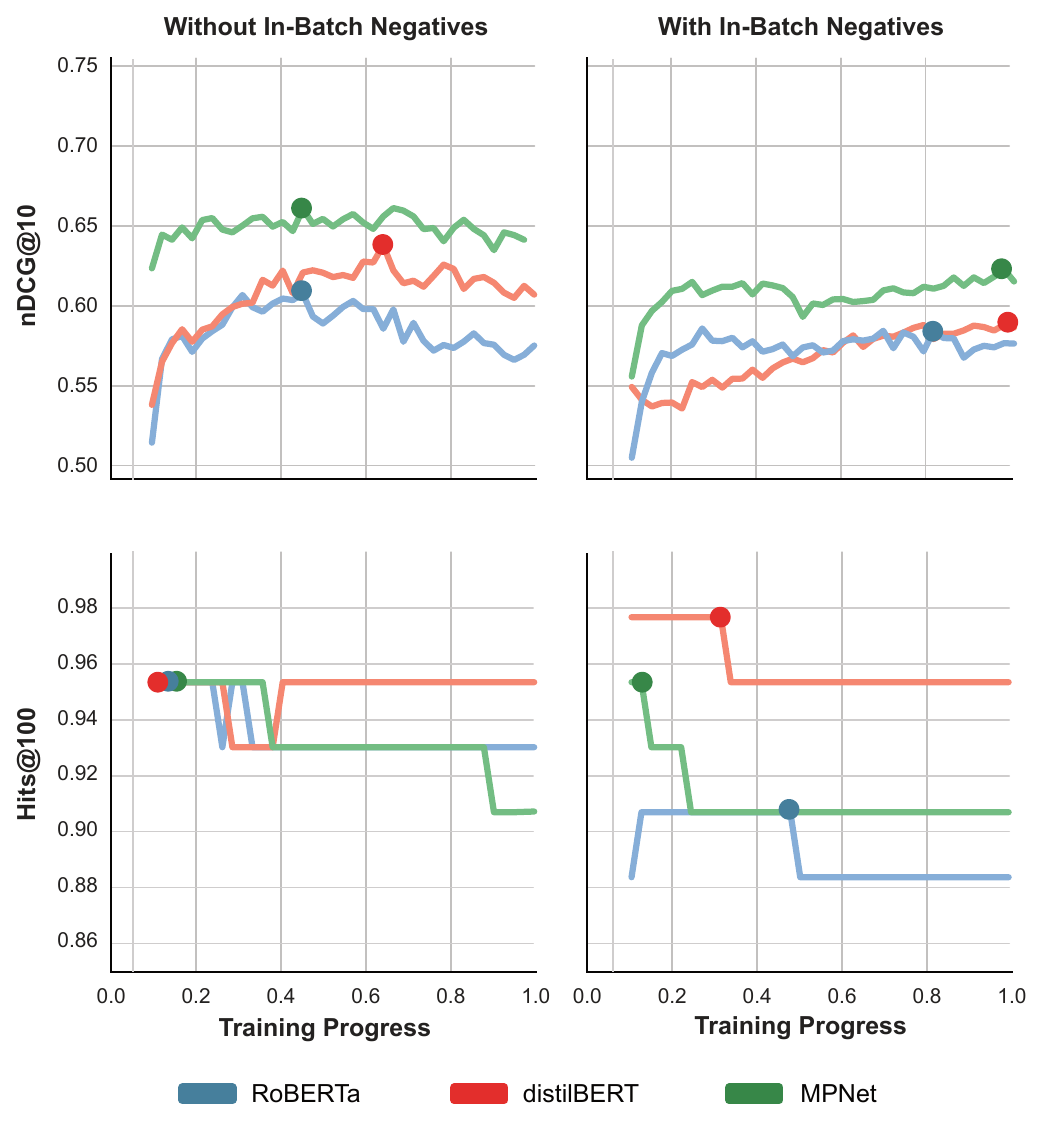}
    \caption{Evolution of nDCG@10 (upper row) and Hits@100 scores (lower row) for adaptive targets on TREC-DL~19, without (left) and with in-batch negatives (left). Plot displays 1 training epoch. Dots mark maxima.}
    \label{fig:adaptive-margins}
\end{figure}

In the second experiment, we examine various configurations of the adaptive target approach. \Cref{fig:adaptive-margins} shows the evaluation scores computed at every 1000\textsuperscript{th} step during the training process. The training uses the entire MSMARCO-Passage collection. Evaluation scores are calculated in the re-ranking setting with a batch size of~$B=128$; similar trends were observed for~$B=64$.

Two key observations can be made:
\begin{enumerate*}[label=(\arabic*)]
\item 
Using adaptive targets is data efficient. The maximum nDCG@10 score is attained after processing approximately half of the data. However, effectiveness declines thereafter, suggesting a divergence from the target task. 
\item 
The use of in-batch negatives produces mixed results in terms of effectiveness. Regarding nDCG@10, across all configurations, the achieved score is lower when using in-batch negatives, and data efficiency is reduced. The maximum score is reached much later in the training process. However, for Hits@100, only RoBERTa does not benefit from the additional information, and the negative impact on data efficiency is diminished.
\end{enumerate*}
Adaptive targets can be effectively employed as a hy\-per\-pa\-ra\-me\-ter-free replacement of the static target variant, yielding comparable evaluation scores. It demonstrates more retrieval-favored characteristics (Hits@100) than the static variant, albeit at the expense of ranking effectiveness (nDCG@10).

\subsection{Distributed Targets}\label{sec:exp-distributed}

\begin{figure}[t]
    \includegraphics[width=\linewidth,trim={0 .7cm 0 .3cm},clip]{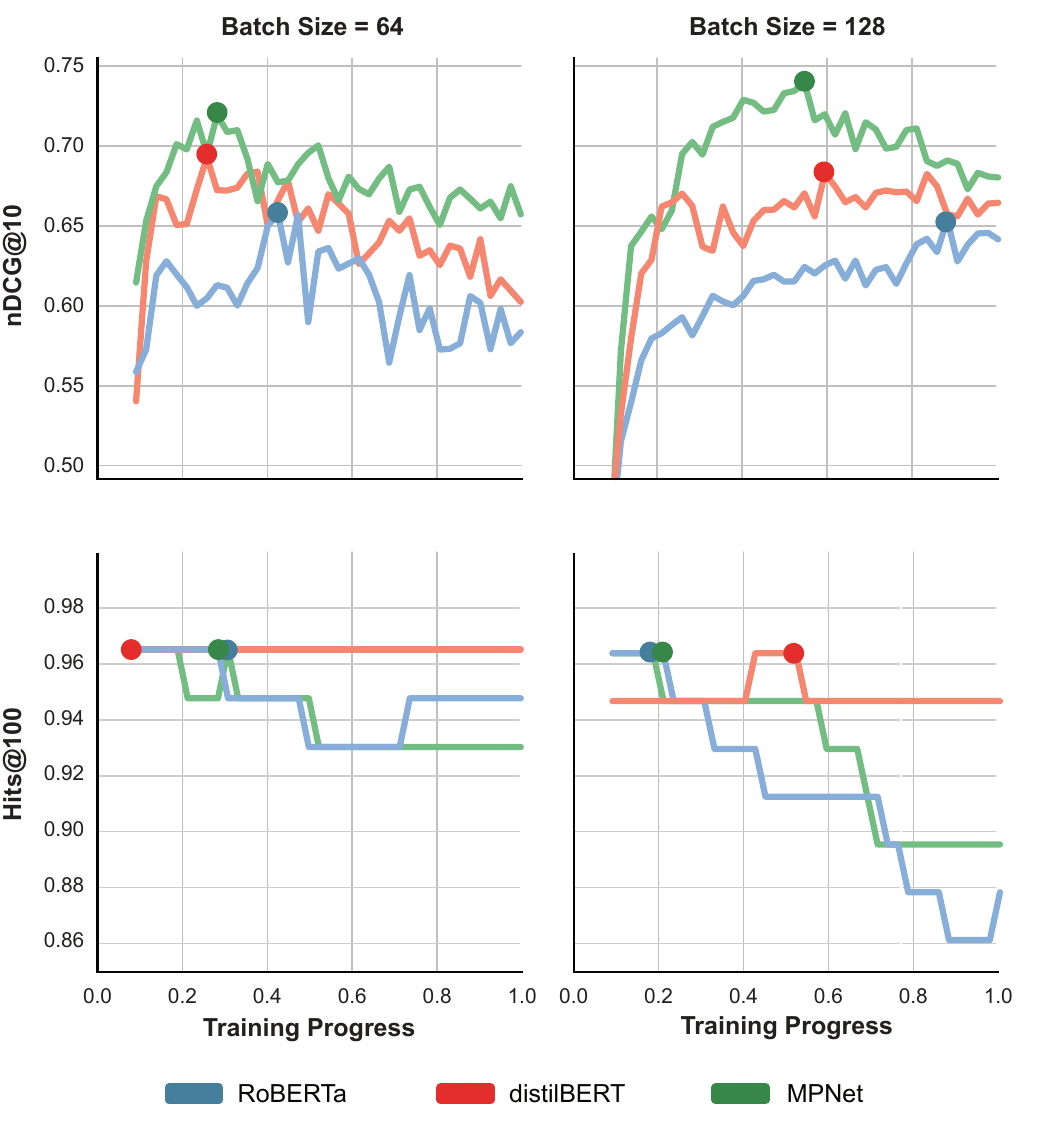}
    \caption{Evolution of nDCG@10 (upper row) and Hits@100 scores (lower row) for distributed targets on TREC-DL~19, with different batch sizes. Plot displays 1 training epoch. Dots mark maxima.}
    \label{fig:distributed-margins}
\end{figure}

In the third experiment, we examine the distributed targets variant. \Cref{fig:distributed-margins} shows evaluation scores calculated in the re-ranked setting for all three pre-trained models. It includes a variation of the batch size, i.e., negative sample count.

Three key observations can be made for the distributed targets variant: 
\begin{enumerate*}[label=(\arabic*)]
\item  
it converges towards the maximum nDCG@10 score even quicker than the adaptive targets variant (see \Cref{fig:adaptive-margins}). Batch size influences convergence, with smaller batches yielding a more data-efficient result.
\item 
The use of in-batch information leads to improved results compared to previous in-batch static or adaptive variants. Specifically, nDCG@10 is highest in the distributed setup.
\item 
Absolute maximum scores are only marginally affected by batch size, with larger batch sizes yielding slightly higher nDCG@10 scores in two out of the three fine-tuned models. 
\end{enumerate*}
Distributed targets also shows to be a robust hy\-per\-pa\-ra\-me\-ter-free replacement for static targets. They exhibit complementary behavior to adaptive targets, showcasing better ranking accuracy~(nDCG@10) while maintaining similar retrieval characteristics~(Hits@100).

\begin{figure}[t]
    \includegraphics[width=\linewidth,trim={0 .7cm 0 .3cm},clip]{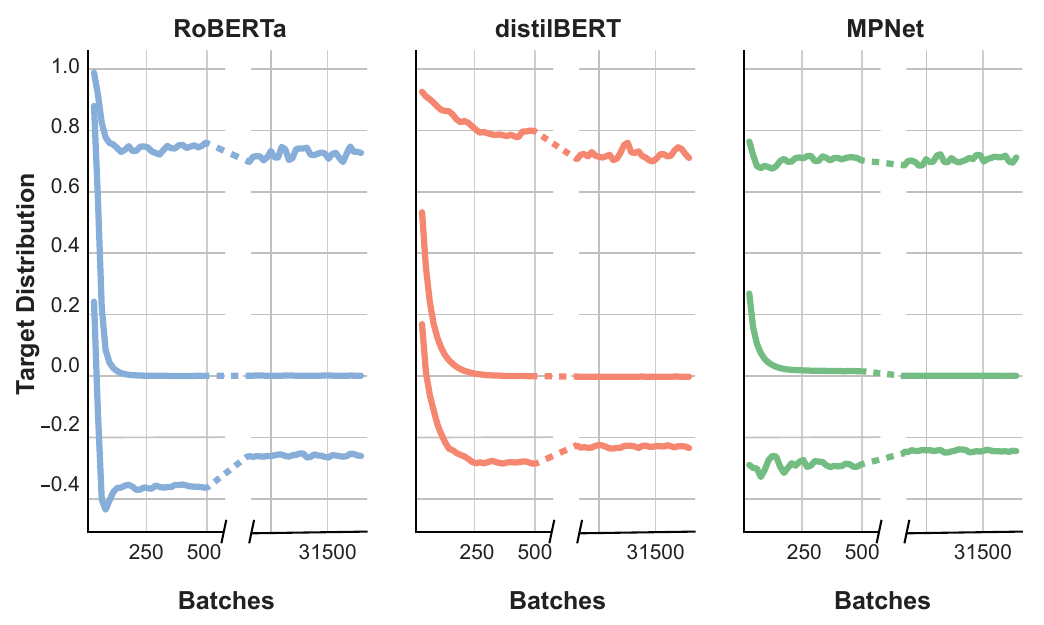}
    \caption{Evolution of mean (middle line), max (upper line) and min (lower line) of target distribution throughout the training process for 1 epoch. Values smoothed by rolling mean over 32 batches. First and last 500 batches shown, in-between omitted (dotted line portion).}
    \label{fig:margin-evolution}
\end{figure}

To contextualize these findings, \Cref{fig:margin-evolution} illustrates the training characteristics of the distributed target variant. It plots the mean, minimum, and maximum values of the $D^+ \times D^-$ similarity matrix throughout the training process. Four observations stand out:
\begin{enumerate*}[label=(\arabic*)]
\item 
The in-batch target distribution rapidly settles to a mean of 0 (rescaled to 0.5 for loss computation), suggesting that non-relevant documents are, on average, orthogonal in latent space to relevant documents. This indicates an ideal class separation in cosine similarity space.
\item 
The distribution features outliers in the positive spectrum, i.e., the maximum value (upper line) is further apart from the mean (middle line) than the minimum value (lower line). Consequently, documents highly similar to the relevant one of a query are less common than dissimilar ones, approaching the expected imbalanced ratio of relevant to irrelevant documents for a query.
\item 
While the initial pre-trained similarity distribution, i.e., the first 100 steps, differs significantly among the three pre-trained models, their fine-tuned distributions are largely identical. This suggests that a robust latent space for retrieval can be generalized independently of pre-training regime and architecture.
\end{enumerate*}

\Cref{tab:loss-example} presents an exemplary excerpt of a training batch, comprising a query with its associated positive and three negative passages (maximum, minimum, and closest to the mean predicted target). Additionally, the table provides the predicted relevance~$\rho$, scaled target~$t$, and instance loss~$l_{ij}^2$ calculated with the fine-tuned distilBERT model.  The theorized loss properties can be observed. The relevance~$\rho$ of documents aligns with their target ordering, i.e., semantic document similarity, showcasing successful self-distillation. Moreover, the less related a negative document is to the positive document, the lower its contribution to the overall loss, showcasing implicit negative mining.

\begin{table}[t]
    \caption{Example of training instance with query ($q$), positive ($d^+$) and negative passages with highest ($d^-_{\max}$), closest to mean ($d^-_\text{mid}$), and lowest ($d^-_{\min}$) respective target score ($t$). Predicted relevance ($\rho$) and resulting instance loss ($l^2_{ij}$) given.}
    \renewcommand{\arraystretch}{.8}
\setlength{\tabcolsep}{5.3pt}
\small
\begin{tabular}{@{}lp{0.62\linewidth}lll@{}}
\toprule 
                                                        & Text                                                                                                           & $\rho$& $t$ & $l_{ij}^2$\\ 
\midrule
$q$                                                     & Where is wild rice grown in USA?                                                     & ---  & ---  & ---  \\
\cmidrule(l){2-5}
$d^+$                                                   
                                                        &  Wild rice (Z. aquatica), also an annual, grows in the Saint Lawrence River, the state of Florida, and on the Atlantic and Gulf coasts of the United States. Texas wild rice is a perennial plant found only in a small area along the {[\dots]} & 0.79 & ---  & ---  \\
\cmidrule(l){2-5}
\multirow{1}{*}{\rotatebox[origin=c]{90}{$d^-_{\max}$\hspace{2.5em}}}        
                                                        & Ivan L. Sander. Northern red oak (Quercus rubra), also known as common red oak, eastern red oak, mountain red oak, and gray oak, is widespread in the East on a variety of soils and topography, often forming pure stands. {[\dots]}    & 0.34 & 0.69 & 0.06 \\
\cmidrule(l){2-5}
\multirow{1}{*}{\rotatebox[origin=c]{90}{$d^-_\text{mid}$\hspace{2.5em}}}    
                                                        & Your answer will appear in the acres field. How to convert acres to ft2 : Use the conversion calculator titled Convert acres to ft2. Enter a value in the acres field and click on the Calculate square feet button. Your answer will {[\dots]}     & 0.06 & 0.51 & 0.05 \\
\cmidrule(l){2-5}
\multirow{1}{*}{\rotatebox[origin=c]{90}{$d^-_{\min}$\hspace{2.5em}}}        
                                                        & 1 Colorado Springs Municipal (Colorado Springs, CO) 2  Right now, 14 airlines operate out of Colorado Springs Municipal. 3  Colorado Springs Municipal offers nonstop flights to 11 cities.  Every week, at least 441 domestic flights and 7 international flights depart {[\dots]}   & 0.12 & 0.46 & 0.04 \\
\bottomrule
\end{tabular}

    \label{tab:loss-example}
\end{table}

\subsection{Comparison of Target Variants}\label{sec:exp-comparison}

To compare all three variants of the proposed loss function, we train a model for each using parameter choices informed by the previous experiments. We use a batch size of $B=128$ and employ exponential learning rate decay~$\gamma = 0.99999$. To prevent the effectiveness degradation observed for adaptive and distributed variants, we implement an early stopping criterion. If no improvement in nDCG@10 score on TREC-DL~19 calculated every 500 batches is achieved for 16 consecutive checks, training is halted.

\begin{table}[t]
    \caption{Comparison of effectiveness per pre-trained model (M), target variant (T), in-batch negatives (IB), and total training time. Maximum per column and model marked bold. $\dagger$~marks runs with statistically equivalent effectiveness compared to maximum. TREC-DL19 grayed out due to use for hyperparameter optimization/early stopping.}
    \label{tab:comparison-variants}
     \centering
    \setlength{\tabcolsep}{4.05pt}
\renewcommand{\arraystretch}{.8}
\small
\begin{tabular}{
    @{}*{3}{l} 
    >{\color{gray}}S[table-format=1.3, detect-weight, table-space-text-post={\pmark}]
    S[table-format=1.3, detect-weight, table-space-text-post={\pmark}]
    >{\color{gray}}S[table-format=1.3, detect-weight, table-space-text-post={\pmark}]
    S[table-format=1.3, detect-weight, table-space-text-post={\pmark}] 
    rS[table-format=1.2, detect-weight, table-space-text-post={M}]@{}
    }
    \toprule
    \multirow{2}{*}{M}                  & 
    \multirow{2}{*}{T}                  & 
    \multirow{2}{*}{IB}                 &
    \multicolumn{2}{c}{nDCG@10}       & 
    \multicolumn{2}{c}{Recall@1000}       &
    \multicolumn{1}{c}{Time}            & 
    \multicolumn{1}{c@{}}{Data Seen}    \\[-.25em]
    
                    \cmidrule(lr){4-5}                      \cmidrule(lr){6-7}         
    &   &           & {\color{black} DL19}     & {DL20}   & {\color{black} DL19}   & {DL20}     &  {(hh:mm)}                  & {(\# Triplets)}        \\
    \midrule
    \multirow{5}{*}{\rotatebox{90}{distilBERT}}
    & S & \xmark    & \bfseries 0.647   & \bfseries 0.632   & \bfseries 0.792 & 0.821 \pmark        & 15:07\textsuperscript{*}           & 5.18{M} \\ 
    & S & \cmark    & 0.574             & 0.569             & 0.786\pmark     & 0.823 \pmark        & 13:30\textsuperscript{*}           & 2.24{M} \\ 
    & A & \xmark    & 0.537             & 0.542             & 0.664           & 0.735               & 2:36\phantom{\textsuperscript{*}}  & 4.93{M} \\ 
    & A & \cmark    & 0.534             & 0.513             & 0.790\pmark     & \bfseries 0.823     & 1:04\phantom{\textsuperscript{*}}  & 1.98{M} \\ 
    & D & \cmark    & 0.639\pmark       & 0.615\pmark       & 0.774\pmark     & 0.809\pmark         & 3:56\phantom{\textsuperscript{*}}  & 7.30{M} \\ 
    \midrule    
    \multirow{5}{*}{\rotatebox{90}{RoBERTa}}
    & S & \xmark    & \bfseries 0.651   & \bfseries 0.625   & 0.771\pmark     &  \bfseries 0.822    & 31:11\textsuperscript{*}           & 6.53{M} \\ 
    & S & \cmark    & 0.578             & 0.510             & 0.767\pmark     &  0.820\pmark        & 31:09\textsuperscript{*}           & 6.53{M} \\ 
    & A & \xmark    & 0.527             & 0.488             & 0.771           &  0.698              & 4:09\phantom{\textsuperscript{*}}  & 3.97{M} \\ 
    & A & \cmark    & 0.572             & 0.492             & \bfseries 0.773 &  0.812\pmark        & 5:20\phantom{\textsuperscript{*}}  & 5.12{M} \\ 
    & D & \cmark    & 0.634\pmark       & 0.594             & 0.766\pmark     &  0.794              & 6:19\phantom{\textsuperscript{*}}  & 6.02{M} \\ 
    \midrule   
    \multirow{5}{*}{\rotatebox{90}{MPNet}}
    & S & \xmark    & \bfseries 0.655   & \bfseries 0.646   & 0.791\pmark     &  0.834\pmark        & 34:55\textsuperscript{*}          &  5.63{M} \\ 
    & S & \cmark    & 0.582             & 0.549             & 0.791\pmark     &  0.835\pmark        & 33:51\textsuperscript{*}          &  4.74{M} \\ 
    & A & \xmark    & 0.561             & 0.573             & 0.655           &  0.733              & 8:10\phantom{\textsuperscript{*}} &  6.21{M} \\ 
    & A & \cmark    & 0.605             & 0.564             & \bfseries 0.817 & \bfseries 0.843     & 7:29\phantom{\textsuperscript{*}} &  5.95{M} \\ 
    & D & \cmark    & 0.635\pmark       & 0.622\pmark       & 0.761           &  0.805              & 5:37\phantom{\textsuperscript{*}} &  4.86{M} \\ 
    \bottomrule \\[-1em]
    \multicolumn{9}{c}{\footnotesize S = static; A = adaptive; D = distributed.\quad \textsuperscript{*}Includes hyperparameter sweep time.} \\
 \end{tabular}

\end{table}

\Cref{tab:comparison-variants} lists the obtained evaluation scores in the full ranking setup on both TREC-DL~19 and TREC-DL~20 for each of the three pre-trained models investigated, each of the three loss variants, and (for static and adaptive targets) with and without in-batch negatives. Statistical equivalence was tested among all approaches per model to determine if the difference of each approach to the respective best per measure and model is meaningful (paired TOST, $\epsilon_L=0.05$, $p < 0.05$, $n = 43$ [TREC-DL~19], $n=54$ [TREC-DL~20], Bonferroni correction applied per test group). Equivalence of approaches is indicated in \Cref{tab:comparison-variants} using a $\dagger$~symbol. Additionally, the table lists the number of training samples seen and the total training time per approach to compare their data/compute efficiency. For all static target variants, this includes compute time spent on hyperparameter search to identify a reasonable target value. Note that TREC-DL~19 is used both to tune hyperparameter values, and to report model effectiveness with. This is deliberate, in order to illustrate a reasonable upper bound on the effectiveness of hyperparameter-based approaches under train/test leakage conditions. TREC-DL~20 presents scores in a neutral setting.

For nDCG@10, the static target approach without in-batch negatives achieves the highest scores across all models and both datasets. However, the distributed targets approach, while scoring slightly lower in all setups is still statistically equivalent, except on TREC-DL~20 with RoBERTa as the base model. This is especially interesting, since the static target approach was privy to test data during hyperparameter optimization, which the distributed approach does not need. For Recall@1000, the adaptive target variant with in-batch negatives achieves the highest scores across all models and both datasets, except for distilBERT on TREC-DL~19, where the static approach without in-batch negatives scores higher. Additionally, in half of the test cases (distilBERT, RoBERTa on TREC-DL~19), the distributed targets are statistically equivalent to the highest score. Therefore, adaptive and distributed targets are suitable hyperparameter-free replacements for tuned static targets for retrieval (Recall@1000) and ranking tasks~(nDCG@10). 

These results also highlight a large difference in training efficiency. Since the static target setup requires hyperparameter tuning, its total training time is between~3x (distilBERT) and~6x (MPNet) higher than the distributed target approach, and between~7x and~15x higher than the adaptive target approach. When comparing efficiency in terms of training triplets seen, two further trends are identified: using in-batch information increases data efficiency, and self-distillation approaches are more data efficient than their static counterparts in all but one case~(distilBERT). In-batch information is also detrimental for static variants, suggesting that our approach to implicit negative mining improves effectiveness. The observed gains in compute- and data-efficiency strengthen the applicability of our proposed approach, as it is not only on par in terms of effectiveness but also does not require large amounts of data or computing time. The inclusion of an additional doc-to-doc similarity computation, making available for training signal in each backpropagation step, could serve as explanation for both the data efficiency, as well as the tendency to overfit in adaptive and distributed margins. The choice of pre-trained model does not yield meaningful differences in training outcome; across all evaluated setups, the absolute scores show no clear difference between the three pre-trained models, and the relative trends concerning the individual loss variants are also stable across all pre-trained models.

\begin{table}[t]
    \caption{Comparison of our self-distillation approach (first block) with teacher distillation (second \& third block) and baseline systems (fourth block). Training setting, loss variant, batch sampling approach, use of in-batch negatives, source, and pre-trained model are listed. Best per block and column marked bold. $\dagger$~marks runs with statistically equivalent effectiveness compared to best reproduced \coloremph{TAS-B run} (\citep{wang:2023}). TREC-DL19 grayed out due to use for hyperparameter optimization/early stopping.}
    \label{tab:comparison-baselines}
    \centering
    \setlength{\tabcolsep}{3.1pt}
\renewcommand{\arraystretch}{.8}
\small
\begin{tabular}{
    @{}
    lcccl 
    >{\color{gray}}S[table-format=1.2, detect-weight,  table-space-text-post={\pmark}] 
    S[table-format=1.2, detect-weight,  table-space-text-post={\pmark}] 
    >{\color{gray}}S[table-format=1.2, detect-weight,  table-space-text-post={\pmark}] 
    S[table-format=1.2, detect-weight,  table-space-text-post={\pmark}]
    r@{}
}
    \toprule
    \multirow{2}{*}{}                 & 
    \multirow{2}{*}{Loss}             & 
    \multirow{2}{*}{\kern-.5em B. Smpl.\kern-.5em}        & 
    \multirow{2}{*}{IB}            & 
    \multirow{2}{*}{Src.}           & 
    \multicolumn{2}{c}{nDCG@10}     & 
    \multicolumn{2}{c}{Recall@1000}     &
    {Data}                       \\            
                                                                             \cmidrule(lr){6-7}                             \cmidrule(lr){8-9}                         
    &             &       &        &                            & {\color{black} DL19}     & {DL20}  & {\color{black} DL19}      & {DL20} & {(\# Tripl.)} \\
    \midrule
    \multirow{3}{*}{\rotatebox{90}{Self-D.}}
    & A                 & Rand. & \xmark & Ours            & 0.54                 & 0.54                  & 0.66                  & 0.74                  &  4.93{M}              \\ 
    & A                 & Rand. & \cmark & Ours            & 0.53                 & 0.51                  & \bfseries 0.79\pmark  & \bfseries 0.82\pmark  &  1.98{M}              \\ 
    & D                 & Rand. & \cmark & Ours            & \bfseries 0.64\pmark & \bfseries 0.62\pmark  & 0.77\pmark            & 0.81\pmark            &  7.30{M}              \\ 
    
    \midrule
    \multirow{8}{*}{\rotatebox{90}{Teacher Distillation\hspace{.25em}}}
    & \multirow{8}{*}{\hspace{.5em}\rotatebox{90}{MarginMSE\quad}}
                    &Rand.  & \xmark &                          & 0.69                 & 0.65                  & 0.77                  & 0.80                  & $\approx$ 22.4{M}   \\
    &               &Rand.  & \cmark & \cite{hofstatter:2021}   & 0.70                 & 0.67                  & 0.79                  & 0.84                  & $\approx$ 22.4{M}   \\
    &               &TAS-B  & \xmark &  (Orig.)                 & 0.69                 & 0.67                  & 0.78                  & 0.83                  & $\approx$ 22.4{M}   \\
    &               &TAS-B  & \cmark &                          & \bfseries 0.71       & \bfseries 0.69        & \bfseries 0.85        & \bfseries 0.87        & $\approx$ 22.4{M}   \\
    
                    \cmidrule(l){3-10}
    &               &Rand.  & \xmark &                          & \bfseries 0.68       & 0.66                  & 0.78                  & 0.82                  & $\approx$ 14.6{M}   \\
    &               &Rand.  & \cmark &  \cite{wang:2023}        & 0.61                 & 0.60                  & 0.66                  & 0.75                  & $\approx$ 14.6{M}   \\
    &   & \coloremph{TAS-B} & \xmark &   (Repro.)\kern-1em      & 0.67                 & \bfseries 0.66        & \bfseries 0.79        & \bfseries 0.83        & $\approx$ 14.6{M}   \\
    &               &TAS-B  & \cmark &                          & 0.66                 & 0.62                  & 0.75                  & 0.79                  & $\approx$ 14.6{M}   \\
    \midrule            
    \multirow{4}{*}{\rotatebox{90}{No Dist.}}
    & BM25          & ---   & ---       & ---                   & 0.50                 & 0.48                  & 0.74                  & 0.81                  & {---}               \\
    & S (T)    & Rand. & \xmark & Ours                        & \bfseries 0.65\pmark & \bfseries 0.63\pmark  & \bfseries 0.79\pmark  & \bfseries 0.82\pmark  &  5.2{M}             \\ 
    & S (T)    & Rand. & \cmark & Ours                        & 0.56                 & 0.53                  & 0.77\pmark            & 0.82\pmark            &  2.2{M}             \\ 
    & S ($\epsilon=1$)    & Rand. & \xmark & \cite{hofstatter:2021}      & .60                  & 0.60                  & 0.71                  & 0.76                  & $\approx$ 22.4{M}   \\
    \bottomrule \\[-.5em]
    \multicolumn{10}{c}{\footnotesize S (T) denotes tuned, S ($\epsilon=1$) denotes untuned static margins.}
\end{tabular}

\end{table}

\subsection{Comparison to TAS-B}\label{sec:exp-baselines}
 
To compare the effectiveness of our proposed approach to established alternatives, \Cref{tab:comparison-baselines} lists nDCG@10 and Recall@1000 scores in the full ranking setup for each of our three loss variants, as well as scores for state-of-the-art bi-encoder models trained via teacher distillation by \citet{hofstatter:2021}. We also include evaluation scores of the reproduction study by \citet{wang:2023}. Both experiments fine-tune the distilBERT model exclusively, therefore we restrict our comparison to distilBERT-based runs. The table is grouped into approaches relying on self-distillation (adaptive and distributed targets), teacher distillation (MarginMSE, original~\cite{hofstatter:2021} and reproduced~\cite{wang:2023}), and no distillation (static targets and BM25). As before, for tuned static targets, TREC-DL~19 is deliberately used both for hyperparameter optimization and to report results.

To compare our approach to teacher distillation, we obtained run files from the reproduction study by \citet{wang:2023}, and tested for statistical equivalence of our approaches against their best-performing run (paired TOST, $\epsilon_L=0.05$, $p < 0.05$, $n = 43$ [TREC-DL~19], $n=54$ [TREC-DL~20], Bonferroni correction applied per test group). Statistical equivalence is indicated in \Cref{tab:comparison-baselines} using a $\dagger$~symbol. No run files were available for the experiments by~\citet{hofstatter:2021}, hence no statistical testing could be carried out here. To compare data efficiency, the table also includes the number of training triplets seen by each model. The early-stopping criterion of \citet{hofstatter:2021} terminates runs after 700-800k~steps at a batch size of~32. This puts a conservative estimate of the data seen during training for their models at least at 22.4M~training triplets. 
\citet{wang:2023} terminate earlier, at \SI{456000}{steps} with the same batch size for their highest-scoring run, i.e., at 14.6M~triplets seen. Their other runs use similar amounts of data. Training time is not comparable due to the different software and hardware setups used between our experiments and prior work. Yet, all teacher distillation approaches require an additional cross-encoder forward pass for the full dataset of \SI{40}{M~triplets} for batch sampling.

\begin{table*}
    \caption{
        Individual zero-shot nDCG@10 across public BEIR datasets and average nDCG@10 scores in-domain (I.-D.) on MSMARCO and zero-shot (Z.-S.). Distributed with and without early-stopping (E.S.). Best per block and column marked bold.}
    \label{tab:beir-benchmark}
    \newcolumntype{R}[2]{%
    >{\adjustbox{angle=#1,lap=\width-(#2)}\bgroup}%
    l%
    <{\egroup}%
}
\newcommand*\rot{\multicolumn{1}{R{22.5}{1em}}}
\robustify\underline
\renewcommand{\arraystretch}{.8}
\setlength{\tabcolsep}{4.73pt}
\small
\centering
\begin{tabular}{
    @{}ll
    *{14}{S[detect-weight,table-format=1.2,round-mode=places,round-precision=2]} 
    p{7pt}
    S[detect-weight,table-format=1.2,round-mode=places,round-precision=2]
    S[detect-weight,table-format=1.2,round-mode=places,round-precision=2]
    @{}
}
\toprule
\multicolumn{2}{@{}l}{System} & 
{TCV} & 
{NFC} & 
{NQ } & 
{HQA} & 
{FQA} & 
{AgA} & 
{T20} & 
{CDS} & 
{QUO} & 
{DBP} & 
{SCD} & 
{FEV} & 
{CFE} & 
{SCF} & 
         &
{I.-D.}  & 
{Z.-S.$\uparrow$}  \\
\midrule
\multirow{7}{*}{\rotatebox[origin=c]{90}{Ours}}
& Adaptive (w/o. IB)            & 0.128    & 0.118    & 0.213    & 0.266    & 0.075    & 0.157    & 0.104    & 0.026    & 0.711    & 0.138    & 0.048    & 0.317    & 0.018    & 0.279    && 0.405    & 0.178    \\ 
& Static (Untuned, w. IB)       & 0.212    & 0.176    & 0.252    & 0.335    & 0.107    & 0.274    & 0.135    & 0.055    & 0.728    & 0.204    & 0.072    & 0.590    & 0.135    & 0.283    && 0.480    & 0.273    \\ 
& Adaptive (w. IB)              & 0.284    &\bf 0.252 & 0.090    & 0.308    & 0.172    & 0.411    & 0.037    & 0.065    & 0.789    & 0.247    & 0.076    & 0.343    & 0.117    &\bf 0.475 && 0.527    & 0.262    \\ 
& Static (Untuned, w/o IB)      & 0.314    & 0.175    & 0.222    & 0.281    & 0.131    & 0.354    & 0.182    & 0.034    & 0.738    & 0.267    & 0.097    & 0.580    & 0.101    & 0.298    && 0.468    & 0.250    \\ 
& Distributed (w/o E.S.)        & 0.316    & 0.214    & 0.351    & 0.422    & 0.174    & 0.337    & 0.173    & 0.061    & 0.803    & 0.254    & 0.102    & 0.585    & 0.132    & 0.402    && .607     & 0.309   \\ 
& Static (Tuned,  w. IB)         & 0.319    & 0.226    & 0.244    & 0.393    & 0.169    &\bf 0.468 & 0.105    & 0.076    & 0.808    &\bf 0.281 & 0.119    & 0.595    &  0.153   & 0.425    && 0.530    & 0.313    \\ 
& Static (Tuned,  w/o IB)        & 0.345    & 0.219    &\bf 0.329 & 0.421    &\bf 0.188 & 0.434    & 0.167    &\bf 0.081 &\bf 0.818 & 0.275    & 0.115    & 0.628    & 0.155    & 0.402    && 0.563    & 0.327    \\ 
& Distributed (w. E.S.)         &\bf 0.395 & 0.228    & 0.283    &\bf 0.438 & 0.165    & 0.329    &\bf 0.205 & 0.075    & 0.800    & 0.274    &\bf 0.127 &\bf 0.678 &\bf 0.156 & 0.441    &&\bf 0.607 &\bf 0.341 \\ 
\midrule
\multirow{4}{*}{\rotatebox[origin=c]{90}{Dns./Lex.}}
&DPR                            & 0.332    & 0.189    &\bf 0.474 & 0.391    & 0.112    & 0.175    & 0.131    & 0.153    & 0.248    & 0.263    & 0.077    & 0.562    & 0.148    & 0.318    && 0.177    & 0.260    \\
&ANCE                           & 0.654    & 0.237    & 0.446    & 0.456    & 0.295    & 0.415    & 0.240    & 0.296    &\bf 0.852 & 0.281    & 0.122    & 0.669    & 0.198    & 0.507    && 0.388    & 0.400    \\
&BM25                           &\bf 0.656 &\bf 0.325 & 0.329    & \bf 0.603& 0.236    & 0.315    &\bf 0.367 & 0.299    & 0.789    & 0.313    &\bf 0.158 &\bf 0.753 & 0.213    &\bf 0.665 && 0.228    & 0.430    \\
&TAS-B                          & 0.481    & 0.319    & 0.463    & 0.584    &\bf 0.300 &\bf 0.429 & 0.162    &\bf 0.314 & 0.835    &\bf 0.384 & 0.149    & 0.700    &\bf 0.228 & 0.643    &&\bf 0.408 &\bf 0.430 \\
\midrule
\multirow{3}{*}{\rotatebox[origin=c]{90}{SOTA}}
&ColBERT                        & 0.677    & 0.305    & 0.524    & 0.593    & 0.317    & 0.233    & 0.202    &\bf 0.350 & 0.854    & 0.392    & 0.145    & 0.771    & 0.184    & 0.671    && 0.425    & 0.440    \\
&Contriever                     & 0.596    & 0.328    & 0.498    & 0.638    & 0.329    & 0.446    & 0.204    & 0.345    &\bf 0.865 & 0.413    &\bf 0.165 & 0.758    &\bf 0.237 & 0.677    && 0.206    & 0.460    \\
&SPLADE                         &\bf 0.711 &\bf 0.345 &\bf 0.544 &\bf 0.686 &\bf 0.351 &\bf 0.521 &\bf 0.243 & 0.341    & 0.814    &\bf 0.442 & 0.159    &\bf 0.796 & 0.228    &\bf 0.699 &&\bf 0.637 &\bf 0.490 \\
\bottomrule
\end{tabular}
\end{table*}   

For nDCG@10, the best self-distillation approach (distributed) scores~0.07 lower than the best original teacher distillation approach (TAS-B batch sampling with in-batch negatives). To the scores attained in the reproduction study, a difference of at most~0.04 can be observed, and the effectiveness is statistically equivalent in all cases. For Recall@1000, a similar trend is observable: compared to the original teacher distillation approach, ours scores lower by~0.06 at most (adaptive with in-batch negatives), yet still statistically equivalent to the reproduction study. This attests to our approach's competitive effectiveness, matching the scores of teacher distillation-based approaches. When comparing the static target approach, we can make two observations: first, tuning the target hyperparameter yields a large increase in Recall@1000, but less so for nDCG@10. Second, it also matches the effectiveness of the teacher-distilled models in the reproduction study, further calling into question if the expense of teacher score inference is worth the marginal increase in effectiveness. Yet, the static targets variant necessitates hyperparameter tuning (see \Cref{tab:comparison-variants}). The data efficiency of the self-distillation approach is between~2x/3x (distributed) and~7x/11x (adaptive) better than the teacher-based approaches.

\subsection{Comparison on BEIR}

We evaluate on the public portion of the BEIR benchmark~\cite{thakur:2021} and comparing to baseline and SOTA scores from the official leaderboard~\cite{kamalloo:2024}. Our comparison systems include: BM25 as a lexical baseline; three dense approaches using in-batch negatives (DPR~\cite{karpukhin:2020}, ANCE~\cite{xiong:2021}, and TAS-B~\cite{hofstatter:2021}); and three SOTA systems (ColBERT~\cite{khattab:2020}, Contriever~\cite{izacard:2022}, and SPLADE~\cite{formal:2021}). Table~\ref{tab:beir-benchmark} presents these results. For fair comparison with TAS-B, we use finetuned distilBERT variants, though MPNet variants showed increased effectiveness ($\approx 0.05$).

The results demonstrate that our proposed approaches achieve competitive performance in in-domain settings while requiring substantially lower computational and data resources than established methods. Specifically the distributed and tuned static margins variants exhibit superior effectiveness, exceeding the score of TAS-B, and nearly being on-par with current SOTA models. Adaptive margins perform slightly worse, but still comptetitive. 

Out-of-domain evaluation reveals patterns similar to in-domain results: distributed and tuned static margins demonstrate superior effectiveness, while adaptive margins perform slightly worse, particularly without in-batch negatives. Parameter tuning for static margins consistently improves effectiveness. However, out-of-domain effectiveness relative to other dense retrievers remains lower than in-domain, confirming the observation of \citet{thakur:2021} that datasets with strong domain shift are problematic for dense retrievers, particularly evident on Touche-2020, FiQA, and CQADupStack.

The distributed approach performs competitively against the best-in-class teacher-based TAS-B method with an average difference of only 0.09 in the zero-shot setting, and exceeding it by 0.2 in the in-domain setting. Overall, it outperforms DPR and closely matching ANCE and TAS-B at  substantially less data and training computation. Early stopping successfully diminishes overfitting, yielding the same in-domain at improved out-of-domain effectiveness. Using MPNet as backbone model, average scores increase by $0.05$ equaling ANCE, and nearly matching TAS-B (paired TOST, $\epsilon_L = 0.05$, $p=0.15$, $n=14$). This positions distributed margins primarily as an highly effective method for retrieval finetuning in in-domain low-resource settings, at diminished but still competitive zero-shot capabilities.

\section{Conclusion}\label{sec:sum}

We introduced self-distillation with adaptive and distributed relevance margins for bi-encoder fine-tuning in retrieval tasks. Our approach introduces two key innovations: a self-distillation mechanism that leverages the model's own similarity assessments, and a distributed target scheme that exploits batch-level information. These eliminate three common requirements in retrieval model training: teacher models, specialized batch sampling, and hyperparameter tuning. Our empirical evaluation demonstrates that distributed targets achieve competitive effectiveness, being statistically equivalent on the same dataset to state-of-the-art teacher distillation methods. As it offers very strong in-domain capabilities, our approach is particularly valuable for environments with limited evaluation data or compute resources. Our results further reveal that traditional teacher distillation offers only marginal improvements over hyperparameter-tuned static margins, questioning the computational cost-benefit ratio of complex teacher architectures. This establishes self-distillation as a practical alternative, especially for purpose-built in-domain models and low-resource settings, since it offers on-par effectiveness with a substantially reduced amount of required data and compute cost. It could find applications in information retrieval beyond bi-encoder training, e.g., extending to pairwise cross-encoder training. While we focus on web text ranking as target task, our proposed approach only relies on a suitable embedding space and could thus be transferred to other modalities, e.g., product, music, or image search.  

\begin{acks}
This work has been partially funded by the German Federal Ministry of Research, Technology, and Space  (BMFTR) under Grant \textnumero~\texttt{01IS24077B}; by the ScaDS.AI Center for Scalable Data Analytics and Artificial Intelligence, funded by the BMFTR and by the Sächsische Staatsministerium für Wissenschaft, Kultur und Tourismus under Grant \textnumero~\texttt{ScaDS.AI}; by a Research Fellowship for Harrisen Scells from the Alexander von Humboldt Foundation; and by the OpenWebSearch.eu project, funded by the European Union (GA~101070014). We thank Shuai Wang for providing us with run files and additional insight about of their reproduction efforts~\cite{wang:2023}.
\end{acks}

\bibliographystyle{ACM-Reference-Format}

\bibliography{ictir25-adaptive-margins}
\end{document}